\documentclass[twocolumn]{aastex701}

\usepackage{graphicx,amsmath,amssymb}
\usepackage[export]{adjustbox}
\usepackage{booktabs}
\usepackage{siunitx}
\usepackage{makecell,caption}
\usepackage{listings}
\usepackage{xcolor}

\newcommand{\kms}{\ensuremath{\mathrm{km\,s}^{-1}}}

\usepackage{txfonts}
\usepackage{bm}
\usepackage[export]{adjustbox}
\usepackage{mathrsfs,amsmath,amssymb}

\def\mnras{Mon. Not. R. Astron. Soc. }

\def\apj{Astrophys. J.}
\def\apjl{Astrophys. J. Lett.}
\def\apjs{Astrophys. J., Suppl. Ser.}

\def\aap{Astron. Astrophys.}

\begin{document}

\title{{A dust-free hierarchically nested supermassive-star model for James Webb Space Telescope Little Red Dots}}

\author[orcid=0000-0003-3993-3249]{Pau Amaro Seoane}
\affiliation{Universitat Polit\`ecnica de Val\`encia, Val\`encia, Spain}
\affiliation{Max Planck Institute for Extraterrestrial Physics, Garching, Germany}
\affiliation{Kavli Institute for Astronomy and Astrophysics at Peking University, Beijing, China}
\email[show]{amaro@upv.es}

\begin{abstract}
Observations by the James Webb Space Telescope reveal a population of high-redshift objects, {the little red dots.} {These sources exhibit} optical reddening alongside blue ultraviolet continua, host broad Balmer lines indicative of active black holes, and lack detectable x-ray emission. Their cosmic abundance at early epochs exceeds standard Eddington-limited growth timescales. Prevailing models invoke specific dust geometries to reconcile these traits. We propose a dust-free physical framework. {We evaluate the concept that the dust-poor subset of little red dots represents the observational manifestation of hierarchically nested supermassive stars.} A primary radiation-dominated envelope traps a nuclear star cluster. Plunging stars drive a magnetic dynamo that inflates the envelope. This produces the red optical continuum. Unobscured infalling stars yield the blue ultraviolet excess. Trapped stellar debris sediments to form a high-density secondary core within this Compton-thick host. This nested topology thermalizes x-rays and powers the broad hydrogen $\alpha$ features. The central seed accretes at the global radiation limit of the host envelope. {This kinetic bypass shortens the assembly timescale from hundreds of millions of years to tens of millions of years.}
\end{abstract}


\section{The early-universe seed paradox}\label{sec.intro}

The infrared resolution of the James Webb Space Telescope isolates a class of high-redshift sources \citep{MattheeEtAl2024}, {the little red dots.} These objects display a v-shaped spectral energy distribution. Their rest-frame optical continua appear red. {This feature has been attributed to dust attenuation.} Their rest-frame ultraviolet slopes appear blue, implying unshielded lines of sight {\citep{MattheeEtAl2024}}.

{The standard inference from these line widths, assuming virialized kinematics, posits the presence of central black holes ranging from }$10^7${ to }$10^8 M_\odot${. This inference acts as a standard interpretation rather than a physical requirement. These objects lack x-ray emission in deep observations \citep{MattheeEtAl2024}.}

The cosmic abundance of little red dots introduces a temporal constraint. A stellar-mass seed requires over $500$ million years of continuous growth at its localized Eddington limit to reach $10^7 M_\odot$. {The age of the universe barely exceeds this assembly time at a redshift of }$5${.} Models attempt to fit these observations into dust geometries. {Phenomenological modeling of little red dots proposes a black hole star configuration \citep{NaiduEtAl2025}.}

{We evaluate a physical framework for the dust-poor subset of this population.} We model little red dots as an intermediate phase of supermassive star evolution. Astronomers postulated supermassive stars to explain the energy output of early quasars \citep{HF63}. These radiation-dominated objects succumb to general relativistic instabilities before nuclear ignition in isolation \citep{Chandra64}.

Supermassive stars assemble at the potential minima of galactic nuclei. Dense nuclear star clusters permeate these regions. The supermassive star acts as a dissipative trap. Background stars traversing the gaseous envelope experience aerodynamic drag. This drag extracts their orbital energy. The injection of kinetic heat arrests the relativistic collapse of the gas cloud \citep{Hara1978}. The phase-space loss-cone dynamics funnel stars into the trapping zone \citep{AS01}.

\begin{figure*}[htbp]
    \centering
    \includegraphics[width=\textwidth]{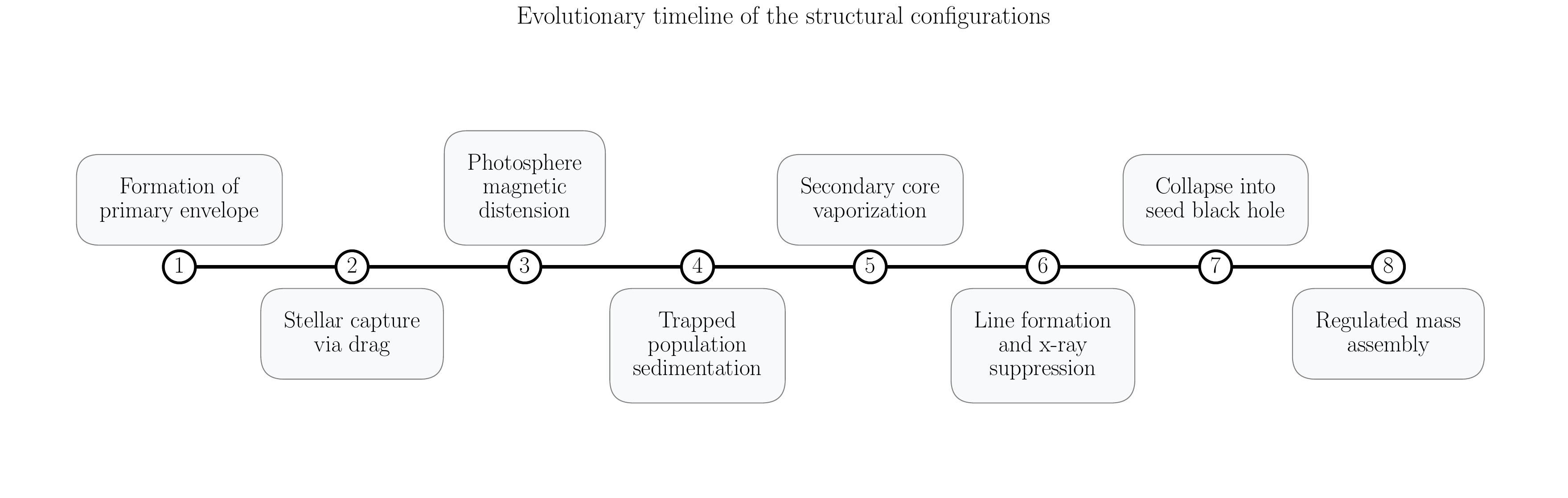}
\caption{{Timeline of the structural evolution. The horizontal axis maps the sequence across }$8${ discrete configurations. The progression spans from the initial gas accumulation to the linear accretion phase.}}
    \label{fig:timeline}
\end{figure*}

{The causal sequence tracks an explicit progression across }$8${ structural configurations. Figure~\ref{fig:timeline} outlines the evolutionary timeline from initial gas accumulation to final black hole formation. First, a primary gas cloud settles at the center of a nuclear star cluster. Second, this extensive cloud acts as a dissipative envelope, capturing orbiting cluster bodies via supersonic aerodynamic drag. Third, this kinetic energy injection fuels an internal magnetic dynamo, inducing geometric distension of the outer photosphere. Fourth, the trapped stellar remnants lose orbital velocity and undergo differential sedimentation toward the potential minimum. Fifth, the localized spatial accumulation triggers a high-velocity collisional cascade, vaporizing the stellar components into a monolithic secondary core. Sixth, the resulting hot core shapes the characteristic broad Balmer line wings and traps high-energy radiation within a Compton-thick barrier. Seventh, the inner core breaches the post-Newtonian stability boundary and undergoes gravitational collapse. Eighth, the resulting seed black hole bypasses localized Eddington accretion limitations, drawing mass continuously under the global radiation limit of the stable surrounding envelope.}

{We evaluate the features of the dust-poor little red dots using this hierarchically nested topology. It yields the v-shaped spectral energy distribution, the broad Balmer kinematics, the lack of x-rays, and the accelerated mass assembly without requiring dust.}

\section{Magnetic distension and the v-shaped spectrum}\label{sec.sed}

A radiation-dominated gas cloud coalesces at the potential minimum of a nuclear star cluster. It functions as a dissipative trap \citep{Hara1978}. Background stars traversing this gaseous medium experience supersonic aerodynamic drag. This interaction extracts their orbital kinetic energy and converts it into thermal heat. {The stellar collision timescale in the core and the subsequent post-Newtonian instability limit the lifetime of the hierarchically nested configuration to a range of }$10^4${ to }$10^5${ years.}

The phase-space loss-cone dynamics of the surrounding nuclear cluster provide a continuous supply of stars \citep{AS01}. The supermassive star accretes angular momentum and deforms into an oblate spheroid. This induces secular resonances that convert regular trajectories into chaotic orbits. Stars on these eccentric paths plunge through the envelope. The supersonic passage leaves behind overlapping convective wakes and drives turbulence.

This mechanical stirring operates as a kinematic dynamo. {The continuous kinetic injection from plunging stars maintains the saturated field confined within the stellar interior against buoyant dissipation. This establishes a time-averaged turbulent state.} We draw upon a general relativistic magnetostatic framework to evaluate this effect \citep{LouYuQing2022}. The transverse magnetic field exerts an outward pressure gradient. {We assume that the stored magnetic energy approaches equipartition with the gravitational binding energy.}

{For a }$10^7 M_\odot${ envelope, the non-thermal support distends the photospheric radius from a standard radiation-dominated value of }$1170 R_\odot${ to }$2 \times 10^6 R_\odot${, yielding an inflation factor of }$1700${. We obtain the unmagnetized radius of }$1170 R_\odot${ from the Lane-Emden solution for a radiation-dominated polytrope evaluated at the post-Newtonian instability limit.}

{The bolometric luminosity of a radiation-supported mass equals its Eddington limit. This geometric expansion of the photospheric radius enforces a drop in the effective surface temperature. The cooling drops the effective temperature to }$3100${ kelvin.}

The magnetically inflated host produces the red rest-frame optical continuum defining the little red dots. It achieves this color without dust. The transverse magnetic field scales dynamically with the local gas density as $\langle B_{\perp}^2 \rangle \propto \rho^2 r^2$, {derived from the conservation of baryon number and Faraday induction \citep{LouYuQing2022}}. Because the field intensity scales with density, it drops from the dense interior to the rarefied outer boundary. This predicts photospheric magnetic field strengths on the order of $10$ to $100$ gauss.

{We establish an order-of-magnitude energy budget. The plunging stars inject kinetic energy at a rate of }$10^{43}${ erg s}$^{-1}${. The magnetic energy content reaches }$10^{54}${ erg. The turbulent dissipation balances the injection rate. The envelope thermal timescale spans }$10^4${ years.}

The spatial inflation sweeps up a larger reservoir of cluster stars. A fraction of the cluster mass remains outside the effective photosphere. The stars residing outside the aerodynamic coupling radius dominate the rest-frame ultraviolet emission. {We estimate that }$10${ percent of the interacting stellar population lies outside the effective photosphere. The ultraviolet radiation from these stars escapes without being reprocessed. The relative weights of the blue and red spectral components change on timescales comparable with the radial oscillations, leading to variability in the v-shaped spectrum.} The composite geometry yields the v-shaped spectral energy distribution, as synthesized in figure \ref{fig:sed_inflation}b.

\subsection{Magnetostatic envelope distension and spectrum synthesis}\label{sec:meth_tov}

We model the thermodynamic state using a general relativistic magnetostatic framework to calculate the radial inflation of the primary gas cloud \citep{LouYuQing2022}. Plunging stars drive a saturated turbulent dynamo. This yields a macroscopic magnetic pressure $\mathcal{P}_{\mathrm{mag}}$ defined by the transverse magnetic energy density $\epsilon_B$.

We couple this non-thermal pressure component into the modified Tolman-Oppenheimer-Volkoff equation,
\begin{equation}
\frac{d}{dr} \left( \mathcal{P}_{\mathrm{tot}} \right) = - 1/2 \left( \rho c^2 + \mathcal{P}_{\mathrm{tot}} + \epsilon_B \right) \frac{d\nu}{dr} - \frac{2 \mathcal{P}_{\mathrm{mag}}}{r} ,
\end{equation}
\noindent where $\mathcal{P}_{\mathrm{tot}} = \mathcal{P}_{\mathrm{therm}} + \mathcal{P}_{\mathrm{mag}}$, the term $\rho c^2$ represents the rest-mass energy density, and $\nu$ defines the metric potential.

The gradient of the magnetic pressure becomes negative in the outer layers. Integrating this equation demonstrates that this outward force exceeds radiation pressure. It inflates the envelope radius by a constant factor relative to an unmagnetized polytrope, {yielding }$R_{\mathrm{mag}} = 1700 R_{\mathrm{st}}${. The subscript standard stands for standard.}
The total luminosity remains Eddington-limited. Applying the Stefan-Boltzmann law dictates that the effective temperature drops to $3100$ K.

\begin{figure*}[htbp]
    \centering
    \includegraphics[width=\textwidth]{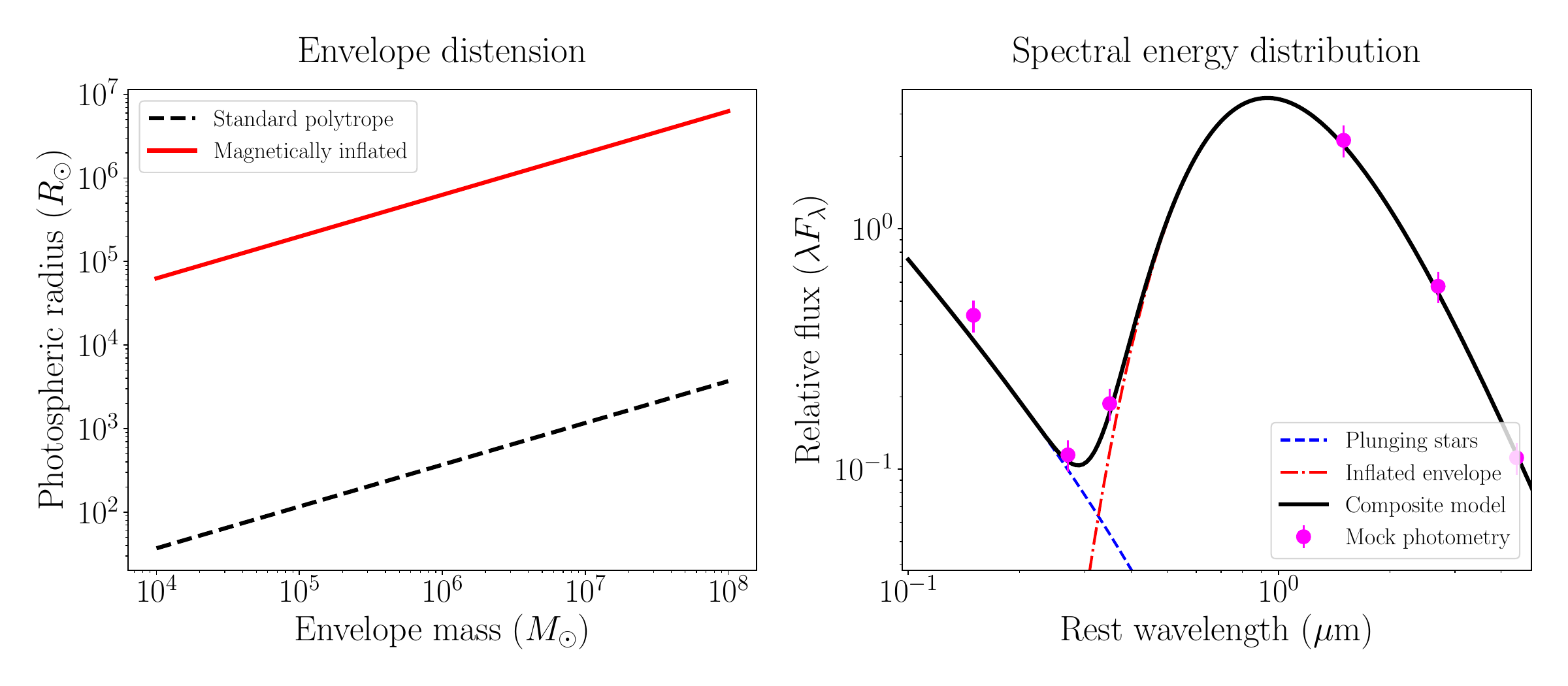}
\caption{Left panel, the mass-radius phase space of the primary supermassive host computed via the modified Tolman-Oppenheimer-Volkoff equations. The dashed blue line traces a standard radiation-dominated polytrope. The solid red curve incorporates the random transverse magnetic field formalism, demonstrating that magnetic pressure inflates the photospheric radius by a factor of $1700$ at fixed mass. Right panel, the synthetic spectral energy distribution generated by this nested geometry. The inflated $3100$ kelvin envelope produces a rising optical continuum. Concurrently, the unshielded cluster stars plunging through the optically thin outer layers emit a blue continuum. The composite spectrum captures qualitative features of the photometry without dust attenuation.}
    \label{fig:sed_inflation}
\end{figure*}

We generate the synthetic spectrum plotted in figure \ref{fig:sed_inflation}b by superimposing this distended thermal blackbody $B_\lambda(T_{\mathrm{eff}})$ with the power-law emission of the plunging stars,
\begin{equation}
F_{\mathrm{total}}(\lambda) = \mathcal{C}_1 \left(\lambda/\lambda_0\right)^{-8/5} \exp(-\lambda/\lambda_{\mathrm{cut}}) + \mathcal{C}_2 B_\lambda(T_{\mathrm{eff}}) ,
\end{equation}

\noindent {where the constants }$\mathcal{C}_1${ and }$\mathcal{C}_2${ act as normalization factors derived from the enclosed unshielded stellar mass and the envelope luminosity. We choose the values of the thermal cutoff }$\lambda_{\mathrm{cut}}${ and the reference wavelength }$\lambda_0${. The index }$-8/5${ models the ultraviolet slope of the unshielded stellar population.}

\section{Nested kinematics and the multi-wavelength deficits}\label{sec.kinematics}

The captured stellar cluster undergoes inward migration. The outward pressure of the tangled magnetic field structurally supports the ionized primary plasma. The individual stars remain globally electrically neutral. The neutral stellar population decouples from the magnetic stress tensor. The stars slip through the supporting magnetic lattice.

We balance the effective gravitational acceleration $|\nabla \Phi_{\mathrm{eff}}|$ against the supersonic aerodynamic ram pressure from the background gas. The terminal slip velocity $v_{\mathrm{slip}}$ of a star of mass $M_\star$ and radius $R_\star$ evaluates to
\begin{equation}
v_{\mathrm{slip}} = \left( 2 M_\star |\nabla \Phi_{\mathrm{eff}}| / (C_D \pi R_\star^2 \rho_{\mathrm{env}}) \right)^{1/2} ,
\end{equation}
\noindent where $C_D = 4$ is the supersonic drag coefficient and $\rho_{\mathrm{env}}$ is the local envelope density. This macroscopic slip dictates that the stellar debris sediments toward the potential minimum. This assembles the secondary core while the primary envelope remains magnetically suspended.

This drift forces a mass concentration at the potential minimum. High-velocity physical impacts grind the discrete stellar remnants into an ionized plasma. The characteristic mass scale decreases to the proton mass. The gravitational P\'eclet number reduces to unity. Stochastic thermal scattering counterbalances the gravitational advection. The debris coalesces into a monolithic secondary core.

\begin{figure*}[htbp]
    \centering
    \includegraphics[width=\textwidth]{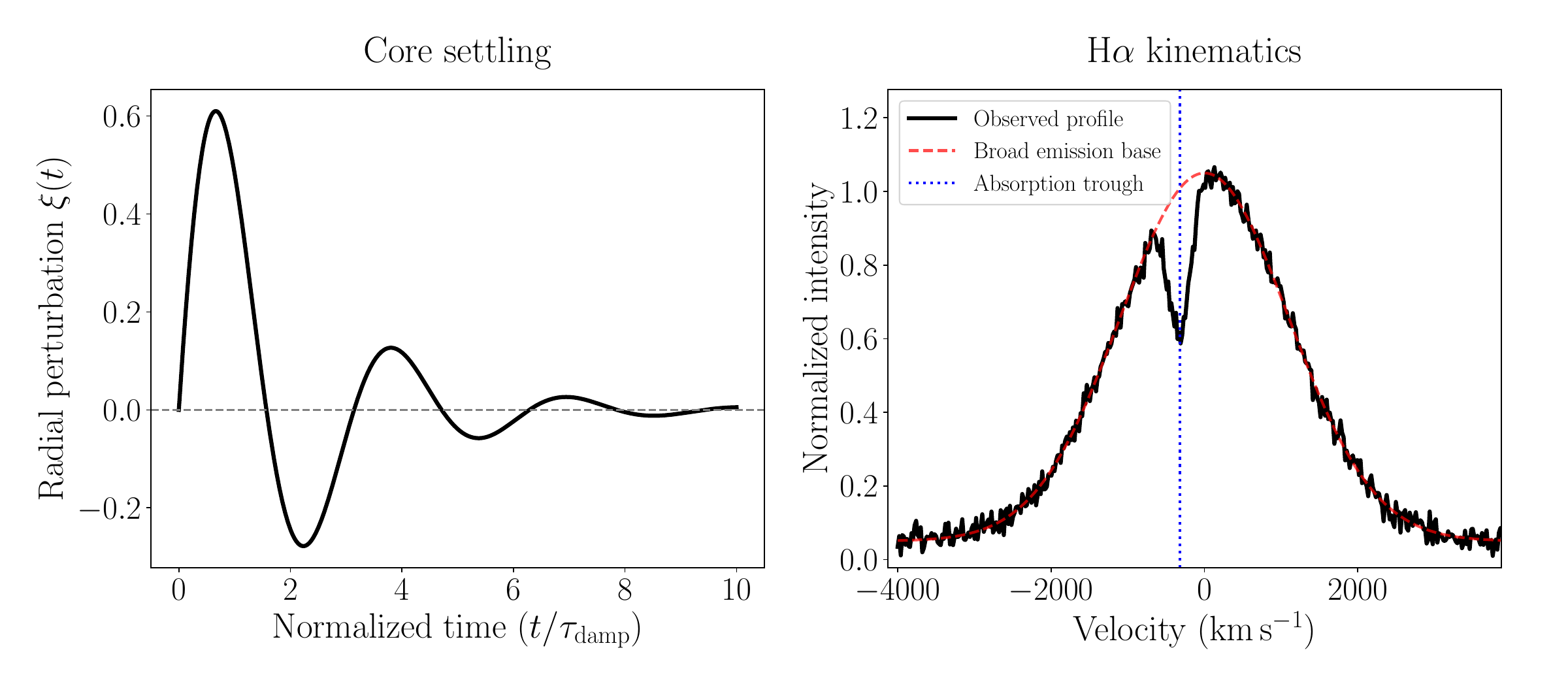}
\caption{Left panel, the dimensionless radial perturbation $\xi(t)$ of the secondary core during the settling phase as a function of normalized time. The damped oscillator formalism captures the thermalization of the stellar debris, describing an initial expansion before settling into exponentially damped quasi-periodic oscillations. Right panel, the resulting synthetic hydrogen $\alpha$ line profile. The virial depths of the secondary core power the broad emission wings. Concurrently, the unvirialized outflowing gas driven by the initial expansion mechanically carves out a narrow absorption trough. The composite synthetic profile captures qualitative features of the complex Balmer profiles.}
    \label{fig:kinematics}
\end{figure*}

The physical environment of this secondary core dictates the emission line kinematics. {We derive the velocity of }$2590 \kms${ from a virial estimate of a }$10^5 M_\odot${ core evaluated at a specified radius of }$130${ astronomical units.} The supersonic flows power the broad Balmer emission wings. The resulting hydrogen $\alpha$ profile exhibits a velocity width of $2590 \kms$, as simulated in figure \ref{fig:kinematics}b, which aligns with the range defining the little red dots \citep{MattheeEtAl2024}.

The macroscopic transverse magnetic field reaches equipartition intensities within the secondary core. The resulting Zeeman separation translates to a velocity shift of $10$ to $100 \kms$. The macroscopic kinematic Doppler broadening obscures this microscopic magnetic splitting. Any Zeeman signature remains undetectable. This preserves the kinematic appearance of the broad spectral features.

Deep spectroscopic surveys reveal multi-peaked Balmer lines with symmetric absorption troughs \citep{MattheeEtAl2024, NaiduEtAl2025}. The vaporized stellar debris originates in a super-virial state. This forces a quasi-adiabatic expansion. {We evaluate the time derivative of the Lagrangian displacement with respect to the systemic velocity of the shell. This quantity determines the relative velocity of the hot core and the cooler envelope. This relative velocity shifts the absorption feature.} Linearizing the restoring force, the temporal evolution of the radial perturbation evaluates to

\begin{equation}
\xi(t) = \mathcal{A}_0 \exp\left(-t/\tau_{\mathrm{damp}}\right) \sin\left( \omega t \right) ,
\end{equation}
\noindent where $\tau_{\mathrm{damp}}$ represents the radiative damping timescale and $\omega$ represents the natural oscillation frequency. The secondary core executes a settling phase, as plotted in figure \ref{fig:kinematics}a. This phase features an initial radial expansion.

The unvirialized gas flows imprint the low-ionization narrow Balmer absorption trough onto the broad virialized emission base. {A blueshifted absorption trough corresponds to a phase where the absorbing material moves toward the observer. A redshifted trough corresponds to a phase where the material contracts. The spatial oscillation remains coherent over the extent of the inner core. We evaluate the fundamental pulsation period of this radiation-dominated structure to determine the variability timescale. The restoring force depends on the gas pressure fraction} $\beta${. The period} $P$ {evaluates to}
\begin{equation}
P = 2\pi \left( \frac{\beta}{2} \frac{G M}{R^3} \right)^{-1/2} ,
\end{equation}
\noindent {where} $M$ {is the mass of the core and} $R$ {is the radius. We evaluate this period for a core of} $10^5 M_\odot$ {settling at} $130$ {astronomical units with} $\beta = 10^{-2}${. The period spans decades. The period decreases to months as the core contracts toward the post-Newtonian instability limit. The radial-oscillation interpretation predicts time evolution of the absorption troughs. The trough velocity correlates with the phase of the radial oscillation and varies on timescales ranging from months to years. This provides a falsifiable observational prediction.}

The hydrogen envelope spherically encases the actively accreting secondary core. The radial column density remains Compton-thick. The surrounding gas mass traps and thermalizes high-energy x-ray photons. Free-free absorption extinguishes low-frequency radio emission.

\section{Bypassing the Eddington limit in the early universe}\label{sec.eddington}

{Classical isolated accretion theory limits the mass evolution of a seed black hole by its localized radiation pressure. The Salpeter timescale equals }$45${ million years. Growing a }$100 M_\odot${ stellar-mass remnant requires over }$500${ million years. The short duty cycle requires a high formation rate. The required seed formation rate correlates with the comoving density of dense nuclear star clusters at a redshift of }$5${.}

Embedding the nascent seed severs the connection between the mass inflow and the localized radiation pressure. {Photon trapping in the optically thick envelope suppresses local feedback. This permits the seed to accrete at the global envelope-regulated rate under saturated convection conditions.} The global mass dictates the maximum stable radiation luminosity.

The global limit transitions the mass evolution from an exponential curve to a rapid linear growth trajectory. The nested black hole achieves the $10^7 M_\odot$ threshold in $67$ million years. Decoupling the accretion limit allows the central engine to accumulate mass in a short timeframe.

\subsection{Super-Eddington growth kinetics}\label{sec:meth_growth}

The mass of the primary host envelope dictates the maximum stable radiation pressure of the composite system. We define this limit as $L_{\mathrm{Edd, env}} = 4\pi G M_{\mathrm{env}} c / \kappa_{\mathrm{es}}$, where $\kappa_{\mathrm{es}}$ represents the electron scattering opacity.

We assume the central seed accretes the gas funneled inward by this global radiation limit. The envelope dictates a constant accretion rate,
\begin{equation}
\dot{M}_\bullet = \frac{M_{\mathrm{env}}}{t_{\mathrm{Sal, adj}}} ,
\end{equation}
\noindent {where the growth time of }$67${ million years represents the adjusted Salpeter timescale }$t_{\mathrm{Sal, adj}}${ for a high-spin geometry.} Decoupling the mass inflow transitions the growth from an exponential to a linear regime. We integrate this constant mass accretion rate to yield the hierarchical growth trajectory,
\begin{equation}
M_\bullet(t) = M_0 + \left( \frac{M_{\mathrm{env}}}{t_{\mathrm{Sal, adj}}} \right) t .
\end{equation}
\noindent {This integration demonstrates that the initial seed matures to the required threshold in }$67${ million years.}

\section{Conclusions}\label{sec.conclusion}

In this work we present the idea that little red dots represent the observational manifestation of hierarchically nested supermassive stars. This configuration explains the multi-wavelength spectrum without invoking dust attenuation. A primary envelope captures background cluster stars via aerodynamic drag, converting orbital kinetic energy into large-scale convection. This mechanical stirring drives an internal magnetic dynamo that amplifies a tangled transverse field. The resulting non-thermal pressure expands the photosphere and lowers the effective surface temperature, establishing the red optical continuum. Concurrently, unshielded stars traversing the outer layers emit the blue ultraviolet excess.

Aerodynamic friction forces the captured stars to sediment toward the central potential minimum, forming a dense secondary core. Inelastic physical collisions vaporize this stellar population into an ionized plasma whose virial velocities generate the broad Balmer emission lines. Because the surrounding envelope is Compton-thick, it thermalizes the central x-ray flux and extinguishes low-frequency radio signatures via free-free absorption. The secondary core eventually undergoes post-Newtonian collapse to form a seed black hole. This seed accretes gas at a rate regulated by the global Eddington limit of the primary envelope rather than its own localized radiation pressure. This mechanical bypass reduces the black hole assembly timescale from hundreds of millions of years to tens of millions of years.

The model yields falsifiable observational predictions. Radial oscillations of the secondary core modulate the spectral energy distribution on timescales of days to weeks. This hydrodynamic process causes the narrow Balmer absorption troughs to shift in velocity according to the internal oscillation phase.

{We note that in this letter we omit exhaustive details to prioritize readability and focus on the cosmological implications of the mechanism. A more detailed analysis of nested supermassive stars will be published elsewhere soon as a companion paper.}

\begin{acknowledgments}
We acknowledge support by China’s National Foreign Expert Program (H).
\end{acknowledgments}


\begin{thebibliography}{}
\expandafter\ifx\csname natexlab\endcsname\relax\def\natexlab#1{#1}\fi
\providecommand{\url}[1]{\href{#1}{#1}}
\providecommand{\dodoi}[1]{doi:~\href{http://doi.org/#1}{\nolinkurl{#1}}}
\providecommand{\doeprint}[1]{\href{http://ascl.net/#1}{\nolinkurl{http://ascl.net/#1}}}
\providecommand{\doarXiv}[1]{\href{https://arxiv.org/abs/#1}{\nolinkurl{https://arxiv.org/abs/#1}}}

\bibitem[{P. {Amaro-Seoane} \& R. {Spurzem}(2001){Amaro-Seoane} \&
  {Spurzem}}]{AS01}
{Amaro-Seoane}, P., \& {Spurzem}, R. 2001, \bibinfo{title}{{The loss-cone
  problem in dense nuclei},} MNRAS, 327, 995

\bibitem[{S. {Chandrasekhar}(1964){Chandrasekhar}}]{Chandra64}
{Chandrasekhar}, S. 1964, \bibinfo{title}{{The Dynamical Instability of Gaseous
  Masses Approaching the Schwarzschild Limit in General Relativity.},} ApJ,
  140, 417

\bibitem[{T. {Hara}(1978){Hara}}]{Hara1978}
{Hara}, T. 1978, \bibinfo{title}{{Evolution of a Super-Massive Star in a Dense
  Stellar System},} Progress of Theoretical Physics, 60, 711,
  \dodoi{10.1143/PTP.60.711}

\bibitem[{F. {Hoyle} \& W.~A. {Fowler}(1963){Hoyle} \& {Fowler}}]{HF63}
{Hoyle}, F., \& {Fowler}, W.~A. 1963, \bibinfo{title}{{On the nature of strong
  radio sources},} MNRAS, 125, 169

\bibitem[{Y.-Q. {Lou} \& J.-Z. {Ma}(2022){Lou} \& {Ma}}]{LouYuQing2022}
{Lou}, Y.-Q., \& {Ma}, J.-Z. 2022, \bibinfo{title}{{Supermassive stars with
  random transverse magnetic fields},} \mnras, 516, 1481,
  \dodoi{10.1093/mnras/stab2631}

\bibitem[{J. {Matthee} {et~al.}(2024){Matthee}, {Naidu}, {Brammer}, {Chisholm},
  {Eilers}, {Goulding}, {Greene}, {Kashino}, {Labbe}, {Lilly}, {Mackenzie},
  {Oesch}, {Weibel}, {Wuyts}, {Xiao}, {Bordoloi}, {Bouwens}, {van Dokkum},
  {Illingworth}, {Kramarenko}, {Maseda}, {Mason}, {Meyer}, {Nelson}, {Reddy},
  {Shivaei}, {Simcoe}, \& {Yue}}]{MattheeEtAl2024}
{Matthee}, J., {Naidu}, R.~P., {Brammer}, G., {et~al.} 2024,
  \bibinfo{title}{{Little Red Dots: An Abundant Population of Faint Active
  Galactic Nuclei at z {\ensuremath{\sim}} 5 Revealed by the EIGER and FRESCO
  JWST Surveys},} \apj, 963, 129, \dodoi{10.3847/1538-4357/ad2345}

\bibitem[{R.~P. {Naidu} {et~al.}(2025){Naidu}, {Matthee}, {Katz}, {de Graaff},
  {Oesch}, {Smith}, {Greene}, {Brammer}, {Weibel}, {Hviding}, {Chisholm},
  {Labb\textbackslash'e}, {Simcoe}, {Witten}, {Atek}, {Baggen}, {Belli},
  {Bezanson}, {Boogaard}, {Bose}, {Covelo-Paz}, {Dayal}, {Fudamoto}, {Furtak},
  {Giovinazzo}, {Goulding}, {Gronke}, {Heintz}, {Hirschmann}, {Illingworth},
  {Inoue}, {Johnson}, {Leja}, {Leonova}, {McConachie}, {Maseda}, {Natarajan},
  {Nelson}, {Setton}, {Shivaei}, {Sobral}, {Stefanon}, {Tacchella}, {Toft},
  {Torralba}, {van Dokkum}, {van der Wel}, {Volonteri}, {Walter}, {Wang}, \&
  {Watson}}]{NaiduEtAl2025}
{Naidu}, R.~P., {Matthee}, J., {Katz}, H., {et~al.} 2025, \bibinfo{title}{{A
  ``Black Hole Star'' Reveals the Remarkable Gas-Enshrouded Hearts of the
  Little Red Dots},} arXiv e-prints, arXiv:2503.16596,
  \dodoi{10.48550/arXiv.2503.16596}

\end{thebibliography}
\end{document}